\documentclass[epj]{svjour}
\usepackage{graphics}
\usepackage{epsfig}

\def\MET{\mbox{${\hbox{$E$\kern-0.6em\lower-.1ex\hbox{/}}}_T$}} 

\def\MP{\mbox{$M_P$}}
\def\mp{\mbox{$M_P$}\ }
\def\TH{\mbox{$T_H$}\ }
\def\mbh{\mbox{$M_{\rm BH}$}\ }     
\def\MBH{\mbox{$M_{\rm BH}$}}         
\def\MET{\mbox{${\hbox{$E$\kern-0.6em\lower-.1ex\hbox{/}}}_T$}} 
\def\met{\mbox{${\hbox{$E$\kern-0.6em\lower-.1ex\hbox{/}}}_T$}\ } 
\def\ifb{fb$^{-1}$}                     

\begin{document}

\title{Black Holes at Future Colliders and in Cosmic Rays}

\author{Greg Landsberg\inst{1}%
\thanks{E-mail: landsberg@hep.brown.edu}%
}                     
\institute{Brown University, Department of Physics, 182 Hope St, Providence, RI 02912, USA}
\date{Received: date / Revised version: date}
%
\abstract{
One of the most dramatic consequences of low-scale ($\sim 1$~TeV) quantum gravity would be copious production of mini black holes at future accelerators and in ultra-high-energy cosmic ray interactions. Hawking radiation of these black holes is constrained mainly to our (3+1)-dimensional world and results in their rapid evaporation. We review selected topics in the mini-black-hole phenomenology, such as production rates at colliders and in cosmic rays, Hawking radiation as a sensitive probe of the dimensionality of extra space, as well as an exciting possibility of finding new physics in the decays of black holes.
} 

\authorrunning{Greg Landsberg}
\titlerunning{Black Holes at Future Colliders and in Cosmic Rays}

\maketitle

\section{Introduction}

The possibility that the universe has more than three spatial dimensions has been pondered upon for well over a century. The realization of this possibility is a keystone idea behind string theory in which extra six or seven {\it compact\/} spatial dimensions are required for the most economical and symmetric formulation of its principles. In particular, string theory requires extra dimensions (ED) to establish its deep connection with the supersymmetry, which leads to the unification of gauge forces. These ED are compactified with the radii of the order of $10^{-32}$~m. 

In a new paradigm~\cite{add}, inspired by string theory, Arkani-Hamed, Dimopoulos, and Dvali (ADD) suggested that several ($n$) of these ED could be as large as $\sim 1$~mm. These {\it large extra dimensions\/} are introduced to solve the hierarchy problem of the standard model (SM) by lowering the Planck scale ($M_{\rm Pl}$) to a TeV energy range. (We further refer to this {\it fundamental\/} Planck scale as $M_P$.) In this new picture, the {\it apparent\/} Planck scale $M_{\rm Pl} = 1/\sqrt{G_N}$ only reflects the strength of gravity from the point of view of a three-dimensional observer. 

Since the original ADD idea appeared in 1998, numerous attempts to find large ED or constrain this model have been carried out. They include measurements of gravity at submillimeter distances~\cite{tabletop}, studies of various astrophysical and cosmological implications of large ED~\cite{astro}, and numerous collider searches for virtual and real graviton effects~\cite{collider}. For a detailed review of the existing constraints and sensitivity of future experiments, see, e.g. \cite{hs}. It is fair to say that the experimental measurements to date have largely disfavored only the case of two or less large ED; for any larger number of them, the lower limit on the fundamental Planck scale is only $\sim 1$~TeV, hardly reaching the natural range of scales expected in the ADD model.

As was pointed out a few years ago~\cite{early}, an exciting consequence of TeV-scale quantum gravity is the possibility of production of black holes (BHs) at the accelerators. Recently, this phenomenon has been quantified for the case of TeV-scale particle collisions~\cite{dl,gt}, resulting in a mesmerizing prediction that future colliders would produce mini black holes at enormous rates (e.g. $\sim 1$~Hz at the LHC for $M_P = 1$~TeV), thus becoming black-hole factories. This observation led to an explosion of the follow-up publications on various properties of mini black holes and made this subject one of the most actively studied aspects of the ED phenomenology.

In this talk we review some of these recent developments. For more extensive review, see, e.g.~\cite{kingman,Cavaglia,SUSYGL}.

\section{Astronomical Black Holes}

While very few people doubt that black holes exist somewhere in the universe, perhaps even abundant, none of the astronomical black hole candidates found so far possess smoking-gun signatures uniquely identifying them as such. Unfortunately, the most prominent feature of a black hole --- the Hawking radiation~\cite{Hawking}~--- has not been observed yet and is not likely to be ever observed by astronomical means. Indeed, even the smallest (and therefore the hottest) astronomical black holes with the mass close to the Chandrasekhar limit~\cite{Chandrasekhar}, have Hawking temperatures of only $\sim 100$~nK, which corresponds to the wavelength of Hawking radiation of $\sim 100$~km, and the total dissipating power of puny $\sim 10^{-28}$~W.\footnote{Note that the event horizon temperature of these black holes is much lower than that of the CMB radiation, so at the present time the BHs grow due to the accretion of relic radiation, rather than evaporate. They will start evaporating when the expanding universe cools down to the temperatures below their Hawking temperature.}

While Hawking radiation would constitute a definite proof of the black hole nature of a compact, massive object, there might be other, indirect means of identifying the existence of the event horizon around such an object~\cite{Narayan}. Probably, the best evidence for the existence of astronomical black holes would come from an observation of gravitational waves created in the collisions of coalescing massive black holes, which LIGO and VIRGO detectors are looking for. However, current sensitivity of these interferometers is still short of the expected signal, even in the optimistic cosmological scenarios. This leaves us with other places to look for black holes that are much smaller and consequently much hotter and easier to detect than their astronomical counterparts.

\section{Properties of Mini Black Holes}

Black holes are well understood general-relativistic
objects when their mass \mbh far exceeds the fundamental (higher
dimensional) Planck mass $\MP \sim 1$~TeV. As \mbh approaches \MP,
the BHs become ``stringy'' and their properties complex. In what 
follows, we will ignore this obstacle\footnote{Some of the properties of the stringy subplanckian ``precursors'' of black holes are discussed in Ref.~\protect\cite{de}.}
and estimate the properties of light BHs by simple semiclassical 
arguments, strictly valid only for $\MBH \gg \MP$. We expect that this 
will be an adequate approximation, since the important experimental
signatures rely on two simple qualitative properties: (i) the
absence of small couplings and (ii) the ``democratic" nature of BH 
decays, both of which may survive as average properties of the 
light descendants of BHs. We will focus on the production and rapid decay of Schwarzschild black holes.

As we expect unknown quantum gravity effects to play an increasingly 
important role for the BH mass approaching the fundamental Planck scale,
following the prescription of Ref.~\cite{dl}, we do not consider BH 
masses below the Planck scale. It is expected that the BH 
production rapidly turns on, once the relevant energy threshold
$\sim\! M_P$ is crossed. At lower energies, we expect BH production
to be exponentially suppressed due to the string excitations or 
other quantum effects. 

Note that the maximum center-of-mass energies accessible at the next generation of particle colliders and in ultrahigh-energy cosmic ray collisions are only a few TeV. Given the current lower constraints on the fundamental Planck scale of $\sim 1$~TeV, the artificial black holes that we might be able to study in the next decade will be barely transplanckian. Hence, the unknown quantum corrections to their GR properties are expected to be large, and therefore we would like to focus on the most robust properties of these mini black holes that are expected to be affected the least by the unknown effects of quantum gravity. Consequently, we do not consider spin and other black hole quantum numbers, as well as grey factors when discussing their production and decay, as their semiclassical values will be modified significantly by these unknown corrections. Having mentioned this important caveat, we refer an interested reader to numerous recent studies of these effects~\cite{BHprop}.

\section{Black Hole Production and Decay}

Consider two partons with the center-of-mass energy $\sqrt{\hat s} =
\MBH$ colliding head-on. Semiclassical reasoning suggests that if the impact parameter of the collision is less than the (higher dimensional) Schwarzschild radius, $R_S$, corresponding to this energy, a BH with the mass \mbh is formed. Therefore the total cross section of black hole production in particle collisions can be estimated from pure geometrical arguments and is of order $\pi R_S^2$. Detailed subsequent studies performed using full GR calculations~\cite{GRcollisions}, as well as other approaches, have confirmed the validity of the geometrical approximation, which we will consequently rely upon.

The Schwarzschild radius of an $(4+n)$-dimensional black hole has been derived in Ref.~\cite{mp}, assuming that all $n$ ED are large ($\gg$ $R_S$).
Using this result, we derive the following parton level BH production cross section~\cite{dl}:
$$
    \sigma(\MBH) \approx \pi R_S^2 = \frac{1}{M_P^2}
    \left[
      \frac{\MBH}{\MP} 
      \left( 
        \frac{8\Gamma\left(\frac{n+3}{2}\right)}{n+2}
      \right)
    \right]^\frac{2}{n+1}.
$$

In order to obtain the production cross section in $pp$ collisions at the LHC, 
we use the parton luminosity approach~\cite{dl,gt,EHLQ}:
$$
    \frac{d\sigma(pp \to \mbox{BH} + X)}{d\MBH} = 
    \frac{dL}{dM_{\rm BH}} \hat{\sigma}(ab \to \mbox{BH})
    \left|_{\hat{s}=M^2_{\rm BH}}\right.,
$$
where the parton luminosity $dL/d\MBH$ is defined as the sum over
all the types of initial partons:
$$
    \frac{dL}{dM_{\rm BH}} = \frac{2\MBH}{s} 
    \sum_{a,b} \int_{M^2_{\rm BH}/s}^1  
    \frac{dx_a}{x_a} f_a(x_a) f_b(\frac{M^2_{\rm BH}}{s x_a}),
$$
and $f_i(x_i)$ are the parton distribution functions. (The dependence of 
the cross section on the choice of the latter is only $\sim 10\%$.)
The total production cross section for $\MBH > M_P$ at the LHC, 
obtained from the above equation, ranges between 15 nb and 1 pb for 
the Planck scale between 1 TeV and 5 TeV, and vteh aries by $\sim 10\%$ 
for $n$ between 2 and 7.

\begin{figure}[tbp]
\begin{center}
\epsfxsize=3.4in
\epsffile{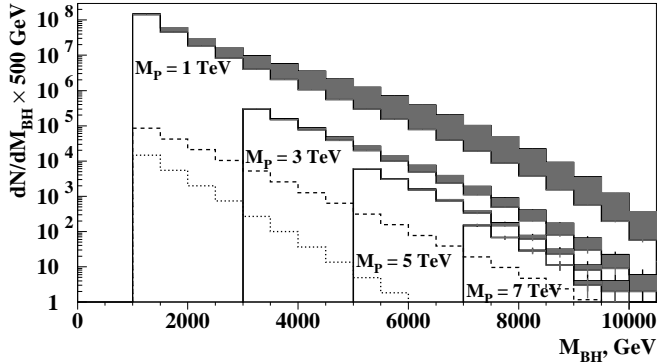}
\caption{Number of BHs produced at the LHC in the electron or photon decay 
channels, with 100~\protect\ifb of integrated luminosity, as a function of the BH 
mass. The shaded regions correspond to the variation in the number of events 
for $n$ between 2 and 7. The dashed line shows total SM background 
(from inclusive $Z(ee)$ and direct photon production). The dotted line 
corresponds  to the $Z(ee)+X$ background alone. From Ref.~\protect\cite{dl}}
\label{nbh}
\end{center}
\end{figure}

Once produced, mini black holes quickly evaporate via Hawking 
radiation~\cite{Hawking} with a characteristic temperature
$$
    T_H = \MP
    \left(
      \frac{\MP}{\MBH}\frac{n+2}{8\Gamma\left(\frac{n+3}{2}\right)}
    \right)^\frac{1}{n+1}\frac{n+1}{4\sqrt{\pi}} = \frac{n+1}{4\pi R_S}
$$
of $\sim 100$~GeV~\cite{dl,gt}. The average multiplicity of particles 
produced in the process of BH evaporation is given by~\cite{dl,gt} and 
is of the order of half-a-dozen for typical BH masses accessible 
at the LHC. Since gravitational coupling is flavor-blind, a BH emits all the 
$\approx 120$ SM particle and antiparticle degrees of freedom with 
roughly equal probability. Accounting for color and spin, we expect 
$\approx 75\%$ of particles produced in BH decays to be quarks and gluons, 
$\approx 10\%$ charged leptons, $\approx 5\%$ neutrinos, and $\approx 5\%$ 
photons or $W/Z$ bosons, each carrying hundreds of GeV of energy. 
Similarly, if there exist new particles with the scale $\sim 100$~GeV, 
they would be produced in the decays of BHs with the probability 
similar to that for the SM species. For example, a sufficiently light 
Higgs boson is expected to be emitted in BH decays with $\sim 1\%$ probability. This has exciting consequences for searches for new physics at the LHC and beyond, as the production cross section for any new particle via
this mechanism is (i) large, and (ii) depends only weakly on particle mass, 
in contrast with the exponentially suppressed direct production mechanism.

A relatively large fraction of prompt and energetic photons, electrons, 
and muons expected in the high-multiplicity BH decays would 
make it possible to select pure samples of BH events, which are also 
easy to trigger on~\cite{dl,gt}. At the same time, only a small fraction 
of particles produced in the BH decays are undetectable gravitons 
and neutrinos, so most of the BH mass is radiated in the form of visible 
energy, making it easy to detect. The mass spectrum of BH events collected at the LHC with 100~fb$^{-1}$ of data and tagged by the presence of an energetic electron or photon among the decay products, along with SM backgrounds, is shown in Fig.~\ref{nbh}~\cite{dl}. It is clear that very clean and large samples of BHs can be produced at the LHC up to Planck scale of $\sim 5$ TeV. Note that the BH discovery potential at the LHC is maximized in the $e/\mu+X$ channels, where background is much smaller than that in the $\gamma+X$ channel (see Fig.~\ref{nbh}). The reach of a simple counting experiment extends up to $\MP \approx 9$ TeV ($n=2$--7), for which one would expect to see a handful of BH events with negligible background. 

A sensitive test of properties of Hawking radiation can be performed by measuring the relationship between the mass of the BH (reconstructed from the total energy of all the decay products) and its Hawking temperature (measured from the energy spectrum of the electron or photon tags). One can use the measured \mbh vs. \TH\ dependence to determine both the fundamental Planck scale \mp and the dimensionality of space $n$. This is a multidimensional equivalent of the Wien's law. Note that the dimensionality of extra space can be determined in a largely model-independent way via taking a logarithm of both parts of the expression for Hawking temperature: $\log(T_H) = -\,\frac{1}{n+1}\log(\MBH) + \mbox{const}$, where the constant does not depend on the BH mass, but only on \mp and on detailed properties of the bulk space, such as the shape of ED~\cite{dl}. Therefore, the slope of a straight-line fit to
the $\log(T_H)$ vs. $\log(\MBH)$ data offers a direct way of determining
the dimensionality of space. The reach of this method at the LHC is discussed in detail in Ref.~\cite{dl}. Note that the determination of the dimensionality of space by this method is fundamentally different from other ways of determining $n$, e.g. by studying a monojet signature or a virtual graviton exchange processes, also predicted by theories with large ED. The latter always depend on the volume of extra space, and therefore cannot offer a direct way of measuring $n$ without making some assumptions about the relative size of various ED. The former, on the other hand, depends only on the area of the event horizon of a black hole, which does not depend on the size of large ($\gg R_S$) ED or their shape.

\section{Discovering New Physics in the Decays of Black Holes}

As was mentioned earlier, new particles with the mass $\sim 100$~GeV would be produced in the process of black hole evaporation with relatively large probability: $\sim 1\%$ times the number of their quantum degrees of freedom. Consequently, it may be advantageous to look for new particles among the decay products of black holes in large samples accessible at the LHC and other future colliders.

As an example~\cite{gl}, we study the discovery potential of the BH sample collected at the LHC for a SM-like Higgs boson with the mass of 130 GeV, which is quite hard to detect via standard means even at the LHC energies. The decay of such a Higgs boson is dominated by the $b\bar b$ final state (57\%).

\begin{figure}[tbp]
\begin{center}
\medskip
\epsfxsize=3.4in
\epsffile{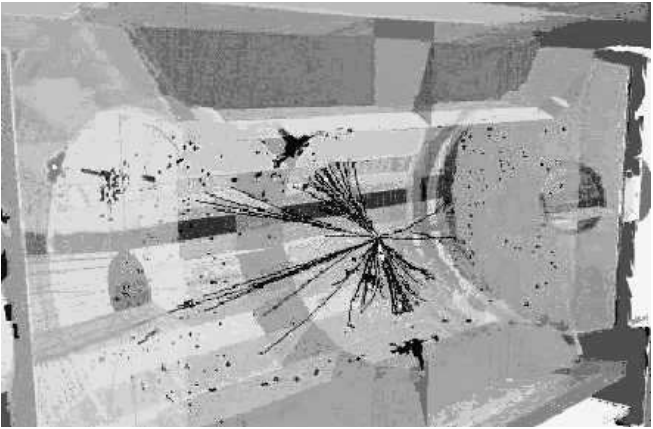}
\caption{A black hole event generated with the TRUENOIR Monte Carlo, followed by the CMS detector simulation code. Courtesy Albert de Roeck and Stephan Wynhoff.}
\label{bh}
\end{center}
\end{figure}

We model the production and decay of the BH with the TRUENOIR Monte Carlo (MC) 
generator~\cite{Snowmass}, which implements a heuristic algorithm to describe a spontaneous decay of a BH. The generator is interfaced with the PYTHIA MC 
program~\cite{PYTHIA} to account for the effects of initial and final state radiation, particle decay, and fragmentation.\footnote{The black hole production and decay has been also implemented in HERWIG~\protect\cite{HERWIG}.} We used a 1\% probability to emit the Higgs particle in the BH decay. We reconstruct final state particles within the acceptance of a typical LHC detector and smear their energies with the expected resolutions. A visualization of a typical black hole event in the CMS detector in shown in Fig.~\ref{bh}.

The simplest way to look for the Higgs boson in the BH decays is to use the 
dijet invariant mass spectrum for all possible combinations of jets found among the final state products. It turns out that Higgs can be firmly established even without $b$-tagging, in spite of all the combinatorics (see Ref.~\cite{gl} for detail). With this method, a $5\sigma$ discovery of the 130 GeV Higgs boson may be possible with ${\cal L} \approx 2$~pb$^{-1}$ (first day), 100~pb$^{-1}$ (first week), 1~fb$^{-1}$ (first month), 10~fb$^{-1}$ (first year), and 100~fb$^{-1}$ (one year at the nominal luminosity) for the fundamental Planck scale of 1, 2, 3, 4, and 5~TeV, respectively, even with incomplete and poorly calibrated  detectors. Thus, for the Planck scale below $\approx 4$~TeV, the integrated luminosity required for discovery via BH decays is significantly lower than that via direct production. 

While this study was done for a particular value of the Higgs boson mass, the dependence of the sensitivity on the Higgs mass in this new approach is small. Moreover, this method is suitable for searches for other new particles with the masses $\sim 100$~GeV, e.g. low-scale supersymmetry. Large sample of black holes accessible at the LHC can be used even to study some of the properties of known particles, see, e.g. Ref.~\cite{uehara}.

\section{Black Holes in Cosmic Rays}

\begin{figure}[tbp]
\begin{center}
\epsfxsize=3in
\epsffile{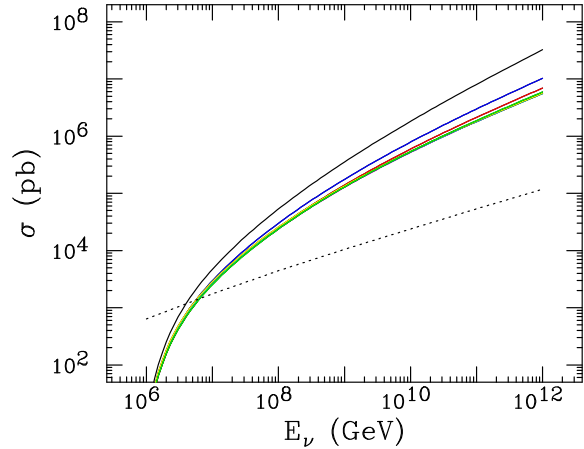}
\caption{Cross sections $\sigma ( \nu N \to {\rm BH})$ for $M_P =
M_{\rm BH}^{\rm min} = 1$~TeV and $n=1, \ldots, 7$. (The last four
curves are virtually indistinguishable.)  The dotted curve is for the
SM process $\nu N \to \ell X$. From Ref.~\protect\cite{Feng}.}
\label{bhcr}
\end{center}
\end{figure}

Recently, it has been suggested that mini black hole production can be also observed in the interactions of ultra-high-energy neutrinos with the Earth or its atmosphere~\cite{Feng}. For neutrino energies above $\sim 10^7$ GeV, the BH production cross section in the $\nu N$ collisions would exceed their SM interaction rate (see Fig.~\ref{bhcr}). Several ways of detecting BH production in neutrino interactions have been proposed~\cite{Feng,CR,afgs,IceCube}, including large-scale ground-based arrays, space probes, and neutrino telescopes as detecting media. 

If the fundamental Planck scale is sufficiently low (1--3 TeV), up to a hundred of BH events could be observed by, e.g. Pierre Auger observatory even before the LHC turns on. These estimates are based on the so-called guaranteed, or cosmogenic neutrino flux~\cite{flux}. In certain cosmological models, this flux could be significantly enhanced by additional sources of neutrino emission, e.g. AGN; in this case even larger event count is possible.

There are two ways to tell the neutrino interaction that results in a BH formation from the standard model processes. The first is based on a particular particle content in the black hole events, and would require good particle identification, perhaps beyond the capabilities of the existing detectors. The second approach is based on the comparison of the event rate for Earth-skimming tau-neutrinos (i.e., those that traverse the Earth crust via a short chord, close to the surface) with that for the quasi-horizontal neutrinos (i.e., those that do not penetrate the Earth, but traverse the atmosphere at a small angle). In the former case, many of the neutrinos would be stopped in the Earth due to the large cross section of BH production, instead of creating atmospheric showers via $\tau$-regeneration in the Earth material. That would suppress the rate of the Earth-skimming neutrino events in a typical ground array detector, such Pierre Auger. At the same time, the rate of the quasi-horizontal events would increase, as the total cross section, which governs this rate, is dominated by the black hole production and therefore is higher than in the SM case. By measuring the ratio of the two rates, it is possible to distinguish the standard model events from the black hole production even with a handful of detected events, despite possible modification of the SM cross section at high energies or uncertainties in the cosmic neutrino flux~\cite{afgs}. Competitive limits on large extra dimensions have been set already by the analysis of the recent AGASA data~\cite{AGASA} on quasi-horizontal showers, which did not exhibit any significant excess above the SM expectations~\cite{afgs}.

\section{Conclusions}

To conclude, black hole production at the LHC and in cosmic rays may be one of the early signatures of TeV-scale quantum gravity. Large samples of black holes accessible by the LHC and the next generation of colliders would allow for precision determination of the parameters of the bulk space and may even result in the discovery of new particles in the black hole evaporation. Limited samples of black hole events may be observed in ultra-high-energy cosmic ray experiments, even before the LHC era.

If large extra dimensions are realized in nature, the production and detailed studies of black holes in the lab are just few years away. That would mark an exciting transition for astroparticle physics: its true unification with cosmology~--- the ``Grand Unification'' to live for.

\section*{Acknowledgments}

I wish to thank the EPS 2003 organizers for their hospitality and an excellent conference. I am grateful to my coauthor on the original black hole paper~\cite{dl}, Savas Dimopoulos, for a number of stimulating discussions and support, and to Jonathan Feng for several conversations on the detection of black holes in cosmic rays. This work has been partially supported by the U.S.~Department of Energy under Grant No. DE-FG02-91ER40688, by the NSF CAREER Award No. PHY-0239367, and by the Alfred P. Sloan Foundation.

\end{document}